\documentclass[]{spie-proceedings-style/spie}  

\usepackage{amsmath,amsfonts,amssymb}
\usepackage{graphicx}
\usepackage[colorlinks=true, allcolors=blue]{hyperref}

\usepackage[textwidth=1.8cm]{todonotes} 
\reversemarginpar
\presetkeys{todonotes}{ size=\small}{} 

\title{Intra-pixel response characterization of a HgCdTe near infrared detector with a pronounced crosshatch pattern}

\author[a]{Charles Shapiro}
\author[a]{Eric Huff}
\author[b]{Roger Smith}
\affil[a]{Jet Propulsion Laboratory, California Institute of Technology, Pasadena, USA}
\affil[b]{Caltech Optical Observatories, California Institute of Technology, Pasadena, USA}

\authorinfo{Send correspondence to C.S.: charles.a.shapiro@jpl.nasa.gov \\ Copyright 2018. All rights reserved.}

\begin{document} 

\newcommand{\reffig}[1]{Fig.~\ref{fig:#1}} 
\newcommand{\refsec}[1]{$\S$\ref{sec:#1}} 
\newcommand{\refeq}[1]{Eq.~\ref{eq:#1}} 
\newcommand{\citeintext}[1]{Ref.~\citenum{#1}}
\newcommand{\um}{\rm \mu m}
\newcommand{\avg}[1]{\left< #1 \right>}

\newcommand{\pasp}{PASP}
\newcommand{\pasa}{PASA}
\newcommand{\procspie}{Proc. SPIE}
\newcommand{\aaps}{AAPS}

\maketitle

\begin{abstract}
The ``crosshatch'' pattern is a recurring ``feature'' of HgCdTe arrays, specifically the Teledyne HAWAII family of near infrared detectors.
It is a fixed pattern of high frequency QE variations along 3 directions generally thought to be related to the crystal structure of HgCdTe.
The pattern is evident in detectors used in WFC3/IR, WISE, JWST, and in candidate detectors for Euclid and WFIRST.
Using undersampled point sources projected onto a HAWAII-2RG detector, we show that the pattern induces photometric variations that are not removed by a flat-field calibration, thus demonstrating that the QE variations occur on scales smaller than the 18 micron pixels.
Our testbed is the Precision Projector Laboratory's astronomical scene generator, which can rapidly characterize the full detector by scanning thousands of undersampled spots.
If not properly calibrated, detectors showing strong crosshatch may induce correlated errors in photometry, astrometry, spectroscopy, and shape measurements.
\end{abstract}


\section{INTRODUCTION}
\label{sec:intro}  

In any image sensor, response to illumination varies from pixel to pixel. 
Such variations may arise from spatial variations in pixel fabrication, substrate thickness, anti-reflective coatings, presence of dust, etc. 
Such variations are typically revealed and characterized by flat field exposures (``flats''), i.e. images of relatively uniform, slowly varying illumination. 
Image processing pipelines typically correct for such variations by dividing astronomical images by the flat;
however, this procedure does not fully remove response variations on scales smaller than a pixel. 
Flat-field calibration only normalizes the mean quantum efficiency (QE) of each pixel to a common level. 
When QE varies within a pixel, measurements of high-contrast images (stars, galaxies, spectral lines) will depend on their location relative to the pixel grid, particularly if the optical point spread function (PSF) is undersampled. 
Such a dependence induces systematic errors in photometry, astrometry, shape measurement, or spectroscopy that may be unacceptable, depending on the desired precision.

\begin{figure} [htp]
   \begin{center}
   \begin{tabular}{cc} 
   \includegraphics[width=4in]{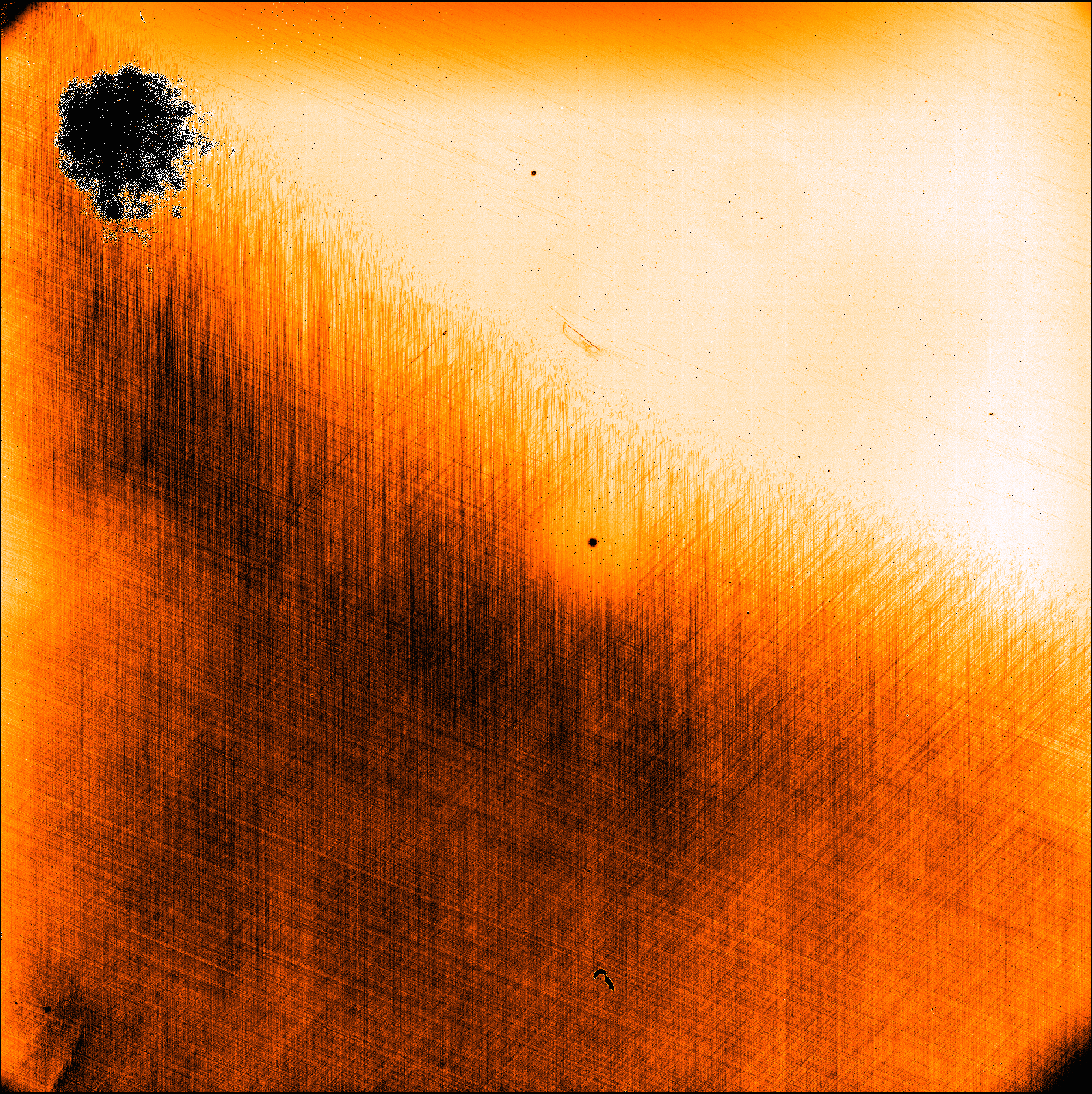} & \includegraphics[width=2.05in]{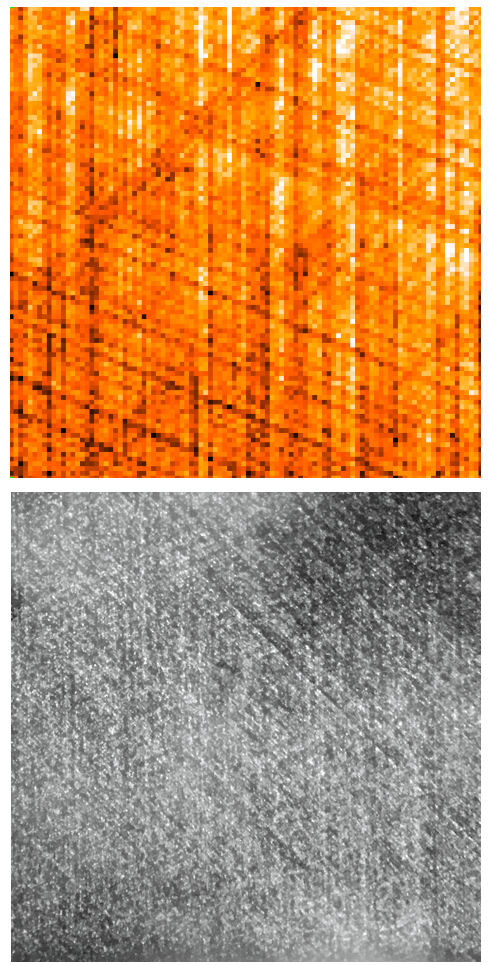}
   \end{tabular}
   \end{center}
   \caption[example] 
   { \label{fig:crosshatch} 
LEFT: Flat-field calibration image (1 $\um$ illumination) from an engineering-grade H2RG detector exhibiting a pronounced crosshatch pattern in the lower left. 
The color scale is stretched to emphasize the pattern; typical pixels have QE ranging from about 1 in the upper right to 0.8 in the lower left. 
A cluster of bad pixels is evident in the upper left corner. 
The illumination is clipped in the corners due to imperfect alignment of one integrating sphere.
RIGHT TOP: Zoomed view of a strongly crosshatched region. 
RIGHT BOTTOM: Photo of a portion of the detector surface taken through an optical microscope.}
   \end{figure}

The HAWAII family of near infrared (NIR) detectors from Teledyne Imaging Sensors\cite{Beletic2008} is widely used in astronomy. 
HAWAII-xRG (or HxRG, where `x' denotes the pixel format) detectors are hybrid Complementary Metal Oxide Semiconductor (CMOS) arrays -- HgCdTe substrates bonded to silicon readout circuits. 
They are popular due in part to their tunable cutoff wavelength (up to 10$\um$), versatile readout capabilities, low dark current, and high QE. 
HxRGs have been employed in instruments for the Hubble Space Telescope (Wide Field Camera 3), the James Webb Space Telescope, and the Wide-field Infrared Survey Explorer. 
They have also been designated for instruments for Euclid, the Wide Field Infrared Survey Telescope (WFIRST), the Thirty Meter Telescope, and many large ground-based observatories.
These detectors have a well-known response variation dubbed the ``crosshatch'' pattern -- a fixed pattern of high frequency QE variations along 3 directions generally thought to be related to the crystal structure of HgCdTe. 
An example of the pattern in shown in \reffig{crosshatch}.

The crosshatch pattern may be more or less pronounced in different regions of a detector or among different detectors; thus, scientific programs may choose to consider it when ranking candidate devices for an instrument.
For instance, the Near Infrared Spectrophotometer (NISP) onboard the Euclid mission will employ H2RG detectors to obtain galaxy color and redshift information needed to map the large scale structure of the Universe. 
The instrument has an error budget of about 1\% relative photometry (RMS), only part of which may be allocated to the detectors. 
NISP has f/10 optics and 18 $\um$ pixels, making it strongly undersampled at wavelengths below 1.8 $\um$, then weakly undersampled up to its 2.3 $\um$ cutoff. 
It is thus sensitive to intra-pixel variations, which may include the crosshatch pattern. 
The primary motivation for this work was to investigate the impact of the crosshatch pattern on NISP photometry.

Testing was carried out at the Precision Projector Laboratory (PPL), a detector characterization facility operated by Jet Propulsion Laboratory\footnote{in collaboration with Caltech Optical Observatories} designed to address detector-related risks to ongoing or proposed space missions through emulation experiments\cite{Shapiro2018}. 
The PPL testbed (henceforth ``the projector'') focuses customized astronomical ÒscenesÓ (e.g. stars, galaxies, spectra) onto detectors at wavelengths from 0.3-2$\um$ and focal ratios of f/8 or slower, spanning many telescopes of interest. 
Aberrations are low over a 40mm square field of view, large enough to cover many large-format image sensors. 
Previous intra-pixel studies \cite{Barron2007,Hardy2014,Hardy2008,Toyozumi2005} have typically used a highly-focused single spot scanner to characterize image sensors. 
The PPL projector's ability to focus tens of thousands of images per scene provides a large multiplex advantage over single-spot scanners. 
By rapidly mapping the whole detector (or a large fraction), we can amass statistics on intra-pixel behavior (small regions may not be representative). 
And by emulating the PSF of a particular instrument such as NISP, we can rapidly characterize the impact of the large scale crosshatch pattern on photometry or other measurements.\footnote{For an alternative approach using fringes, see e.g. \citeintext{Crouzier2014,Crouzier2016}}

\section{Experimental Setup}

   \begin{figure} [ht]
   \begin{center}
   \begin{tabular}{c} 
   \includegraphics[width=4.5in]{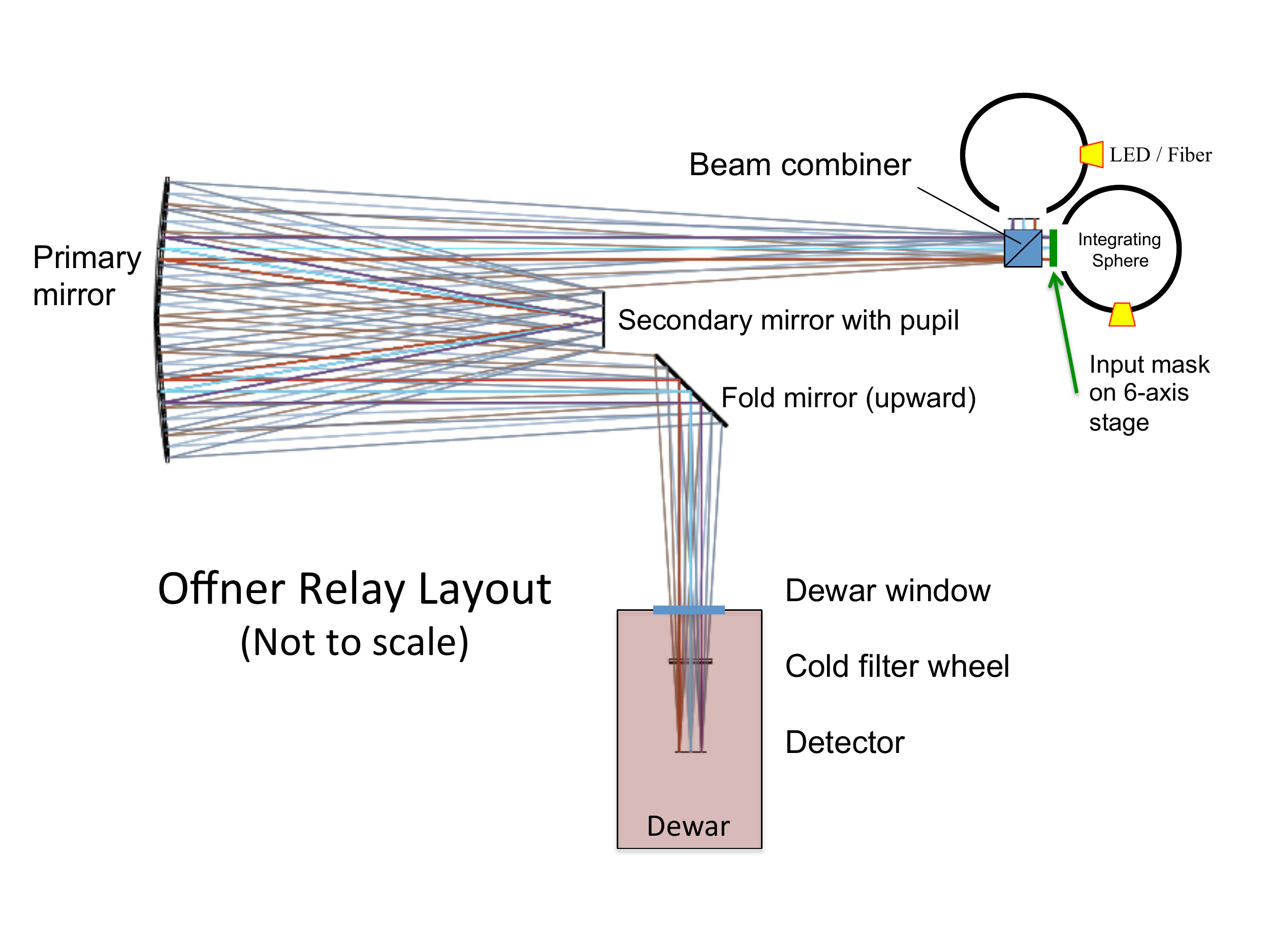}
   \end{tabular}
   \end{center}
   \caption[example] 
   { \label{fig:offner} 
Layout of the ``projector'' testbed. 
The Offner relay focuses a 1:1 image of an illuminated target mask (e.g. a grid of spots) or flat field onto detectors with low aberrations over a 40x40mm field of view. 
The mask, image orientation, f-number, and illumination are easily adjusted to provide a range of stimuli. 
See \citeintext{Shapiro2018} for more details.}
   \end{figure} 

The projector testbed is based on an Offner relay design as illustrated in \reffig{offner}. 
Integrating spheres fed by LEDs or by fiber optics from a quartz-tungsten-halogen lamp illuminate a target mask (glass or quartz) containing a ``scene'' etched in a chrome coating. 
A beam combiner cube allows us to easily switch between two scenes -- such as a mask and a flat field -- or superimpose them. 
A 1:1 image of the mask is focused onto the detector with excellent optical performance over a wide area due the cancellation of aberrations by the symmetry of the input and output paths. 
When optics are properly aligned, the Strehl ratio is greater than 90\%. 
The focal length is fixed at 750 mm, and focal ratios from f/8 and slower are selectable by changing pupil stops at the secondary mirror. 
A fold mirror reflects the beam upward through a turntable where we mount the detector being tested; this enables 360 degree rotation of the detector about the optical axis with constant gravity vector. 
  For infrared detectors, which are sensitive to thermal radiation from the room-temperature testbed ($>1.3\um$), selecting an illumination wavelength requires us to match a cold filter in the cryostat (dewar) to the wavelength of the LED or a filter on the lamp.

The projector's optical PSF is well-approximated by an Airy disk with full-width-half-maximum given by ${\rm FWHM}=1.03 \lambda F$, where $\lambda$ is the illumination wavelength and $F$ is the f-number. PSF size is thus controlled via the f-number and wavelength to obtain various degrees of sampling from a detector with fixed pixel pitch. 
Using masks containing point sources (``pinholes'' a few microns wide), spot widths smaller than 1 pixel can be produced for most wavelengths and pixel sizes of interest. 
When the detector response is factored into the PSF, the total PSF is slightly blurred by lateral charge diffusion. 
Due to air turbulence (lab seeing), the PSF is also slightly broadened by image motions of about 1 $\um$ rms.

The NIR detector tested in this experiment (H2RG 18546) was loaned to PPL by the Euclid detector working group. 
It is an engineering-grade Teledyne H2RG-18 NIR detector ($2048^2$ pixels; 18$\um$ pitch) with a cutoff wavelength of 2.3$\um$. 
The detector was operated at a temperature of 95K with a room-temperature Leach controller clocked at 166kHz in 32-channel readout mode. 
Image data is acquired via ``up-the-ramp'' sampling, in which the pixels are reset and then their voltages read at regular intervals while photo-charge integrates. 
To minimize data reduction of thousands of images, we typically acquire 4 samples of the full detector area at a rate of 3s per frame. 
We discard the 1st sample to avoid contamination by settling effects just after reset. 
Fluxes are inferred by fitting a quadratic function to the remaining samples for each pixel and extracting slopes after correcting for nonlinearity. We limit fluence to about 70,000$e^-$/pixel (about 1/2 full well) to ensure the quadratic model is an acceptable approximation. 
We apply a mean conversion gain of 2.7 e-/ADU to all pixels and find that the detector has a median read noise of 10e- per raw frame (14e- for correlated double sampling).

A flat-field image shown in \reffig{crosshatch} (LEFT) clearly distinguishes a ``strong'' crosshatch region in the image lower left from a ``weak'' crosshatch region in the upper right. 
The weak crosshatch region, which looks smooth by comparison but which does exhibit the crosshatch pattern, is more representative of flight-grade detectors. 
The partitioning of the detector into these two regions provides a convenient way to analyze the impact of the pattern on measurements that emulate Euclid photometry.  
The detector also has a pattern of hot pixels resembling the crosshatch pattern, which is most noticeable when the detector is above operating temperature (see \refsec{hotpixel}).

We use the projector to focus a grid of about 18,000 spots (3$\um$ pinhole apertures in the mask) with a uniform spacing of 274.5$\um$ (15.25 pixels) onto the detector. 
Selecting our cold Y-band filter (0.9 -- 1.07$\um$) and an f/11 focal ratio produces a diffraction limited PSF with a full-width-half-maximum of about 11$\um$, broadened to about 14$\um$ (0.78 pixels) by lateral charge diffusion and seeing. The PSF flux is thus strongly concentrated within the area of a single 18$\um$ pixel, while the 3$\um$ aperture profiles remain unresolved. 
The grid is scanned laterally in steps of 6$\um$ (1/3 pixel), approximating the smallest scale resolved by the optical PSF. 
The spatial Nyquist rate set by the PSF is approximately $2/{\rm FWHM}\sim 1/(7\um)$, so smaller steps do not provide additional information. 
Focus and scanning are achieved by moving the mask on a 6-axis stage with automated stepper motors.

\section{Power spectrum of the crosshatch pattern}

\begin{figure} [htp]
   \begin{center}
   \begin{tabular}{c} 
   \includegraphics[width=6in]{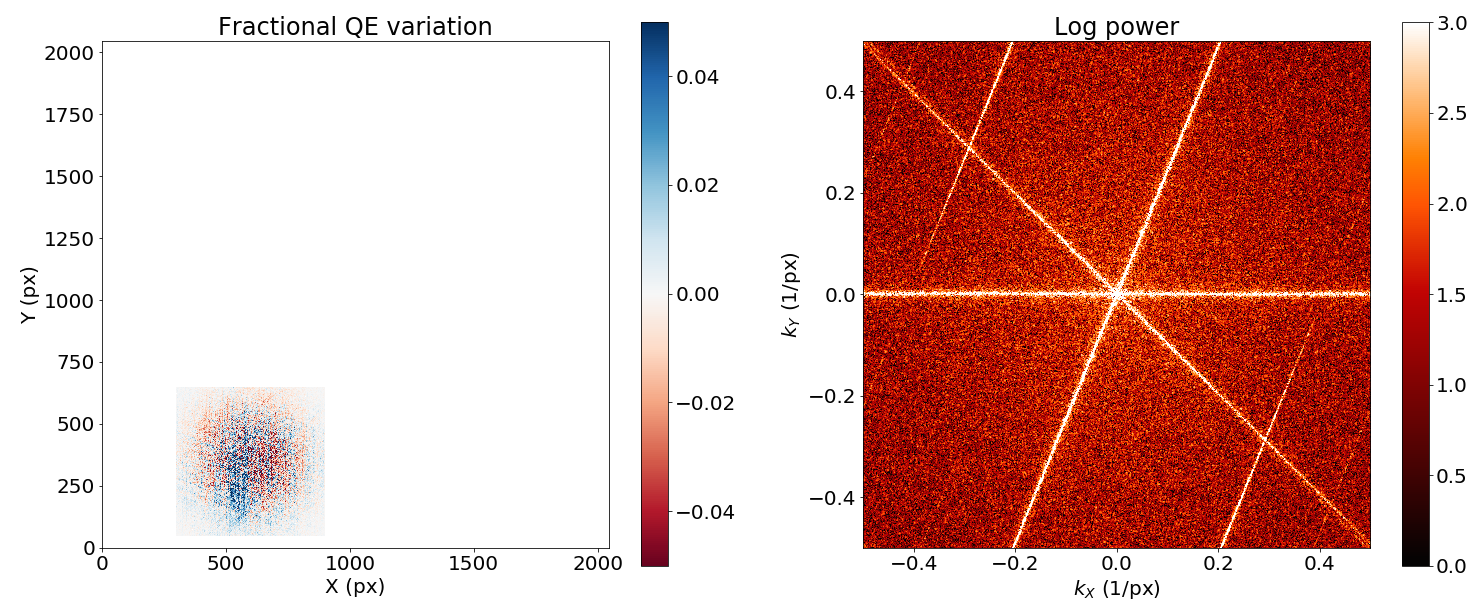} \\
   \includegraphics[width=6in]{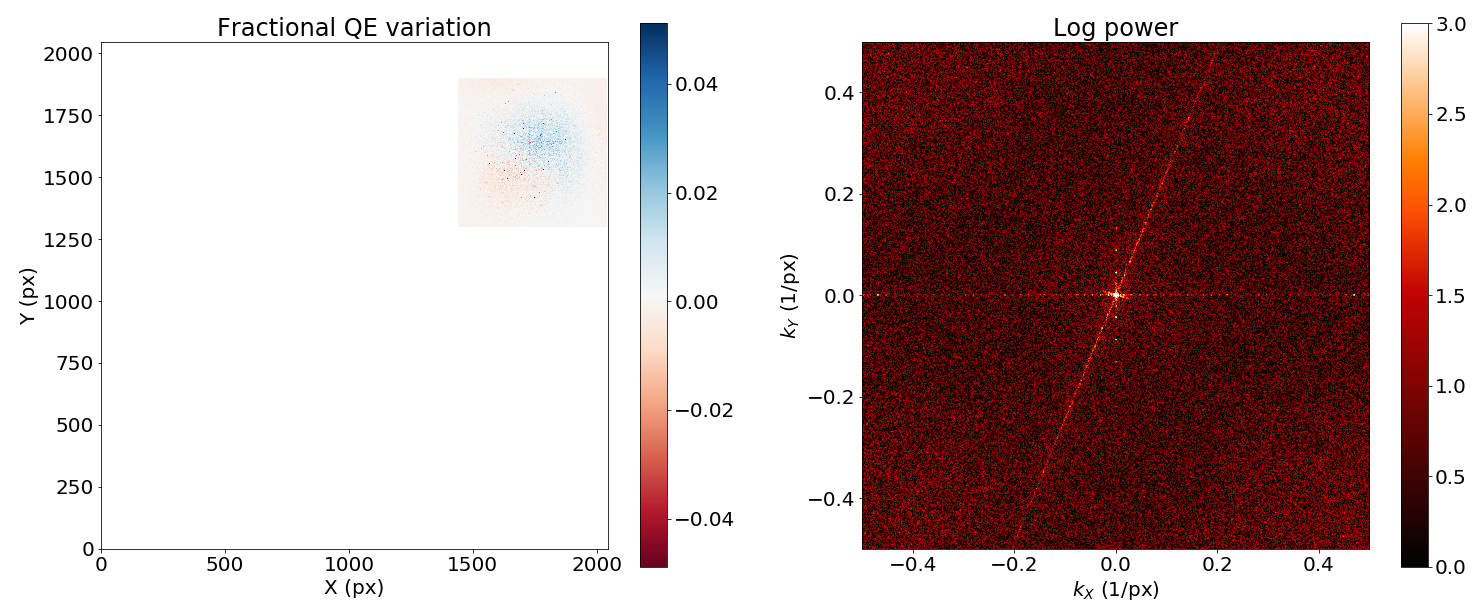}
   \end{tabular}
   \end{center}
   \caption[example] 
   { \label{fig:crosshatch-power} 
LEFT: Flat field calibration image, masked to isolate part of the strong (top) or weak (bottom) crosshatch region. 
The color scale denotes QE fluctuations relative to the mean of the subregion.
A gaussian window function is applied to reduce edge effects. 
RIGHT: 2D power spectra (squared Fourier transform) of each subregion. 
The prominent lines of power have wave-vectors pointing along 0, 68, and -45 degrees, approximately.}
   \end{figure}

Before delving into intra-pixel measurements, we begin with a closer look at the flat image in \reffig{crosshatch} (LEFT). 
We select two 600x600 pixel$^2$ subregions of the image which are wholly contained within the strong or weak crosshatched detector regions. 
For each subregion, we compute the QE variations relative to the mean,
$ \Delta{\rm QE} \equiv {\rm QE}/\avg{\rm QE}-1 $,
creating the maps shown on the left in \reffig{crosshatch-power}. 
The subregions are multiplied by gaussian windows with $\sigma=150$ pixels, and finally the power spectra (2D Fourier series amplitudes) are computed and shown on the right. 
The gaussian windows attenuate ringing in the Fourier transform due to sharp edges of the subregions.

The power spectra reaffirm what we can see by eye, that the strong crosshatch region has much more structure at high frequencies. 
Furthermore, it is composed of three discrete components defined by linear (1D) power spikes along three directions (0, 68, and -45 degrees, approximately).\footnote{0 and -45 degrees are interesting angles.  If the pattern is truly related to the HgCdTe crystal lattice, they imply that the pixel grid was deliberately aligned to the lattice during fabrication.} 
If the crosshatch along one direction were a sinewave with a single spatial frequency, we'd see a concentrated node of power in the spectrum. 
A line of power, instead, means that the pattern varies along a single direction but on a range of scales. 
Since the lines reach the edge of the plot, there is power at least up to the Nyquist frequency (1/2px). 
Interestingly, if we trace the 68 degree line from the center towards higher frequencies, it first reaches the Nyquist frequency in the $y$ (vertical) direction and then switches sign in $k_y$, continuing from about $|k_x|=0.2$/px on the other side of the plot. 
We interpret this as the familiar aliasing or ``folding'' effect caused by frequencies higher than Nyquist. 
On close inspection of the figure, one can even see the line fold again around the $k_x$ axis, appearing faintly in the upper left and lower right corners.
Power in the 0 and -45 degree lines will also be aliased, but in these cases, the power folds over onto and is indistinguishable from the power in the lower frequency modes along those same directions.

The apparent presence of power at spatial frequencies beyond Nyquist is highly suggestive of intra-pixel QE variations, but it is technically not a smoking gun. 
By simulating an otherwise ideal detector with some diagonal lines of dead pixels -- i.e. response variations that are precisely one pixel wide and aligned with the pixel grid -- one can easily generate a power spectrum with spikes extending to high frequencies. Similarly, the flat field images from the H2RG could conceivably arise from piece-wise constant ``top-hat'' response variations that are one pixel wide and which align with the pixel grid. 
In that case, response variations could indeed be corrected in astronomical images by simply dividing by the flat. 
Admittedly, that scenario seems somewhat contrived, but we cannot rule it out empirically using the flat, which only shows us the average response of each pixel. 
High contrast images are needed to conclusively demonstrate intra-pixel variations; nevertheless, the power spectra of flat field calibration images provide a simple diagnostic for the presence of the crosshatch pattern.

\section{Demonstration of Photometry Degradation}

When a detector is properly calibrated, photometric measurements should not depend on an image's position on the detector. 
Without explicitly mapping the QE on sub-pixel scales (as we do in \refsec{wiener}), we can assess the impact of the crosshatch pattern by quantifying how photometry responds to small displacements of the image with standard calibrations applied. 
The spot grid was scanned through 3 pixels in 1/3 pixel steps in the vertical (column) direction to avoid crossing channel boundaries, and 10 exposures were taken at each grid position. 
We can thus compare variations in spot fluence measurements for a sequence of images at a fixed position to a sequence of images at different positions. 
In the absence of intra-pixel QE variations, the scatters in fluence from these ``moving'' and ``fixed'' datasets should match.

We measure spot fluences via aperture photometry -- unweighted sums in 6 pixel diameter apertures. 
These are computed using the SExtractor\cite{Bertin1996} {\tt FLUX\_APER} measurement applied to post-calibrated images. 
The calibrations applied are bias drift correction (using H2RG reference pixels), dark and background subtraction, nonlinearity correction, and most importantly, the flat-field response. 
Aperture photometry is insensitive to crosstalk due to inter-pixel capacitance (IPC), which partially redistributes the signal in a pixel to its neighbor pixels. 
Nevertheless, we apply an IPC correction to the flat (in addition to the aforementioned corrections) since we want to divide out high-frequency QE variations that themselves would be smeared out by IPC in the flat exposure. 
We assume that the effect of IPC is to convolve images by a $3\times3$ matrix, $(I+K)$, where
\begin{equation}
I=
\begin{pmatrix}
0 & 0 & 0 \\ 
0 & 1 & 0 \\
0 & 0 & 0 \\
\end{pmatrix},\:
K=
\begin{pmatrix}
0 & 0.007 & 0 \\ 
0.009 & -0.032 & 0.009 \\
0 & 0.007 & 0 \\
\end{pmatrix} .
\label{eqn:IPCkernel}
\end{equation}
The average kernel $K$ was measured by Teledyne Imaging Sensors and confirmed at PPL by inspecting the signals adjacent to hot pixels. 
We thus convolve the flat calibration image by $(I-K)$, which cancels out the IPC convolution $(I+K)$ to first order in $K$. 
Bad pixels were identified by looking for outliers in dark frames and in the linearity calibration, and spots near bad pixels were excluded from analysis.

   \begin{figure} [htp]
   \begin{center}
   \begin{tabular}{c} 
   \includegraphics[width=7in]{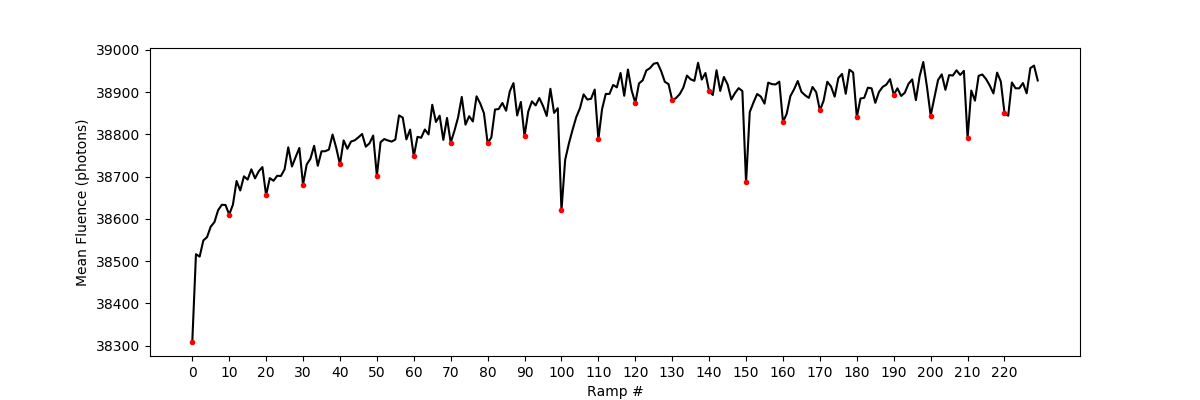}
   \end{tabular}
   \end{center}
   \caption[example] 
   { \label{fig:burn-in} 
Mean photometry of $\sim$16,000 spots over a sequence of exposures in a scan. 
The lamp shutter is closed just before exposure multiples of 10 (marked by red dots) while the image is repositioned. 
The corresponding sharp dips in signal are due to ``burn-in,'' a slight loss in detector sensitivity at the onset of illumination.}
   \end{figure} 

\reffig{burn-in} shows the mean spot fluence over sequential exposures in the scan. 
The signal dips by up to 1\% on exposure multiples of 10, which correspond to exposures taken just after the the grid was repositioned with the shutter closed for about 3 seconds. 
This is evidence of ``burn-in,'' an effect in HxRG detectors wherein pixels gain sensitivity after illumination or, equivalently, lose sensitivity after sitting in darkness. 
A related effect is persistence -- a delayed signal which appears in previously illuminated pixels. 
Both effects are related to charge traps in the HgCdTe, with burn-in arising from photocharges becoming trapped and persistence arising from photocharges being released.\cite{Regan2012}  To mitigate these effects, we discard data from the beginning of each sequence, thus allowing the photometric measurements to settle. 
We then detrend the remaining data by dividing the individual spot fluence sequences by relative variations in the mean, which also removes fluctuations due to illumination instability.

\begin{figure} [htp]
   \begin{center}
   \begin{tabular}{c} 
   \includegraphics[width=5in]{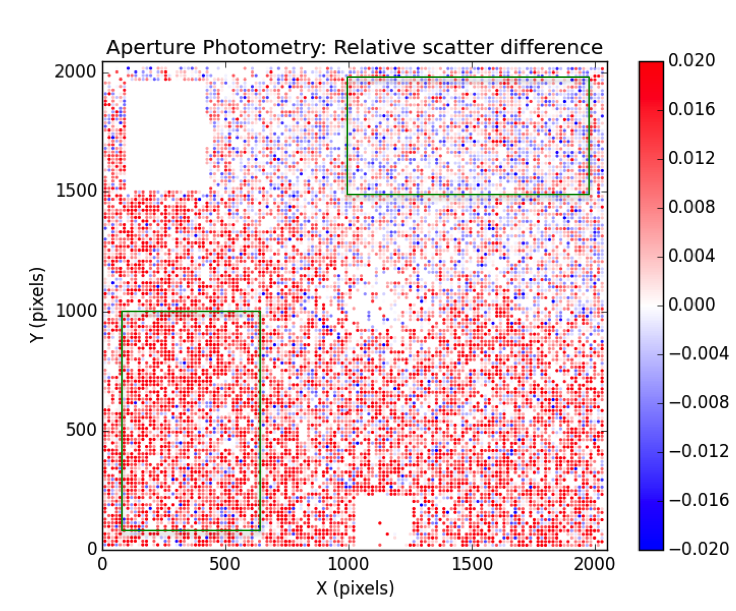}
   \end{tabular}
   \end{center}
   \caption[example] 
   { \label{fig:photometry-demo} 
We measure aperture photometry for a grid of spots, comparing a sequence of stationary images to a sequence with small offsets.  
We map the difference in photometric scatter between the 2 sequences ($\Delta$ in \refeq{diffscatter}) and mask spots near bad pixels.  
Moving the grid around induces percent-level errors in the strong crosshatch region, even when flat-field variations have been removed.  
Averaged over the two subregions denoted by green boxes, we find $\overline\Delta=0.01\pm0.0005$ in the strong crosshatch region and $\overline\Delta=0.0002\pm0.0006$ in the weak region.
}
   \end{figure}

Having calibrated and detrended the fluence data, we choose exposures \#211-219 for the fixed sequence and \#129,139,...,219 for the moving sequence. 
For each spot, we compute the mean fluence $\mu$ and the standard deviation $\sigma$ for the two sequences. 
We then compute
\begin{equation} \label{eq:diffscatter}
\Delta \equiv \frac{\sigma_{\rm moving}}{\mu_{\rm moving}} - \frac{\sigma_{\rm fixed}}{\mu_{\rm fixed}}
\end{equation}
for each spot to see whether the photometry is noisier as a spot is observed in slightly different locations. 
The ``fixed'' term will be dominated by shot noise and read noise while the ``moving'' term will contain extra contributions from any spatially dependent detector effects. 
A map of $\Delta$ is shown in \reffig{photometry-demo}, with each spot displayed at its starting position (grid spacing is 15.25 pixels and the image was scanned by only 3 pixels).
The map shows a clear demarcation between the strong crosshatch and weak crosshatch regions. Moving the image by small amounts induces additional photometric scatter for spots in the strong region, up to about 2\%. 
The effect in the weak region is much less pronounced. 
When we average the map over the two subregions denoted by green boxes in \reffig{photometry-demo}, we find $\overline\Delta=0.01\pm0.0005$ in the strong region and $\overline\Delta=0.0002\pm0.0006$ in the weak region.
If we skip the flat-fielding step before measuring photometry, we find $\overline\Delta=0.016\pm0.0005$ in the strong region and $\overline\Delta=0.0009\pm0.0007$ in the weak region; thus, detrending the pixel-scale variations does mitigate photometric errors but not completely.
Although the average spot in the weak region exhibits no additional photometric scatter, averaging dilutes effects that may be present in a few isolated subregions.

This demonstration reveals that the crosshatch pattern is associated with photometric variations not removed by the usual flat-field calibration procedure. 
We thus infer the presence of QE variations on sub-pixel scales. 
If the crosshatch pattern in the flat image were instead caused by charge redistribution (cf.~ the ``tree-ring'' effect in Dark Energy Survey CCDs\cite{Plazas2014}) and not varying QE, aperture photometry would be unaffected -- signal would only be shifted among pixels without changing the sum. 
Regardless of the mechanism, we have shown that Euclid and other scientific programs can mitigate photometric errors by selecting detectors with weaker crosshatch, as diagnosed by e.g. the power spectra of flat field images. 
By approximating the NISP PSF and analyzing large sections of the detector, we are able to make this determination using measurements that are directly relevant to Euclid photometry requirements.
Note that for the same detector properties, the magnitude of the photometric error will decrease when PSF size is increased, and thus will depend on wavelength in a diffraction limited instrument even if intra-pixel response is invariant with wavelength.

\section{Intra-pixel response mapping} \label{sec:wiener}

Mapping the QE for the entire detector down to sub-pixel scales is achieved by tracking aperture photometry as a function of position.  
We raster scan the spot grid image  10 times in 1/3 pixel (6 $\um$) steps over a 15.25 pixel range in X and Y (the grid spacing) so that each pixel is covered by at least one spot (with some overlap).  
In order to mitigate trends due to charge-trapping, the scan pattern is interlaced so that a spot in a given exposure is at least 5 pixels away from any spot in the previous exposure (spot FWHM $\sim 14\um=0.78$ pixels).  
We simply discard a spot from the data if it ever comes within 5 pixels of a known bad pixel; about 16,000 of 18,000 are kept.  
The motorized stage for the mask can only be commanded with about 1$\um$ precision, but we can measure the actual trajectory much more precisely by averaging the centroids of the $\sim$16,000 useable spots. 
The resulting dataset is an irregularly sampled 2D grid of photometry measurements, which we interpolate onto a regular grid via Wiener filtering, described below.  As before, calibrations including dividing out the flat-field image are applied before aperture photometry is measured. 


\subsection{Theory}
We treat the measured spot photometry as noisy, irregularly spaced samples from an underlying ``true'' pattern of spot photometry. 
We implement the algorithm for optimal interpolation between noisy, irregularly-sampled data described in \citeintext{1992ApJ...398..169R} to combine our multi-epoch offset spot photometry data into a single set of measurements of a common grid.

Let the vector of recorded spot intensities (fluences) $\boldsymbol{p}$ be a linear combination of signal $\boldsymbol{s}$ and noise $\boldsymbol{n}$:
\begin{align}
  \boldsymbol{p} = \boldsymbol{s}+\boldsymbol{n} \;,
\end{align}
which have respective signal and noise covariance matrices $S_{ij}$ and $N_{ij}$ between intensity measurements $i$ and $j$.  
Since spot profiles are non-overlapping and measurements come from separate exposures, we treat the noise matrix as diagonal:
\begin{align}
N_{ij} = \sigma_i^2 \delta^{\rm D}_{ij}
\end{align}
where $\sigma_i^2$ is the variance of a single spot (shot noise plus read noise) and $\delta^{\rm D}_{ij}$ is the Kronecker delta function.  We model the signal covariance as
\begin{align}
S_{ij} = P(\boldsymbol{x}_i - \boldsymbol{x}_j)
\end{align}
where $P(\boldsymbol{x})$ is a spot profile as a function of detector position $\boldsymbol{x}$.  If we think of the spot photometry as a point measurement of the PSF-smoothed response surface of the detector, then nearby measurements are correlated due to the finite size of the PSF.  We assume that signal and noise are uncorrelated.

Indexing interpolation points with $\mu$, then the  maximum-likelihood estimate for the interpolated photometry is
\begin{align}
\hat{\boldsymbol{p}}_\mu = S_{\mu i}\left[S_{ij}+N_{ij}\right]^{-1} (\boldsymbol{p}_j - \langle\boldsymbol{p}\rangle) + \langle\boldsymbol{p}\rangle
\label{eqn:wiener}
\end{align}
where $\langle\boldsymbol{p}\rangle$ is the mean spot intensity. This expression is effectively Wiener filtering the spot maps onto a common grid. 
Equation~\ref{eqn:wiener} can also be used to forecast the impact of large-angle subpixel correlations on high-level science outcomes, such as galaxy clustering applications.

\subsection{Maps and discussion of features}

\begin{figure} [htp]
  \begin{center}
   \includegraphics[width=\textwidth]{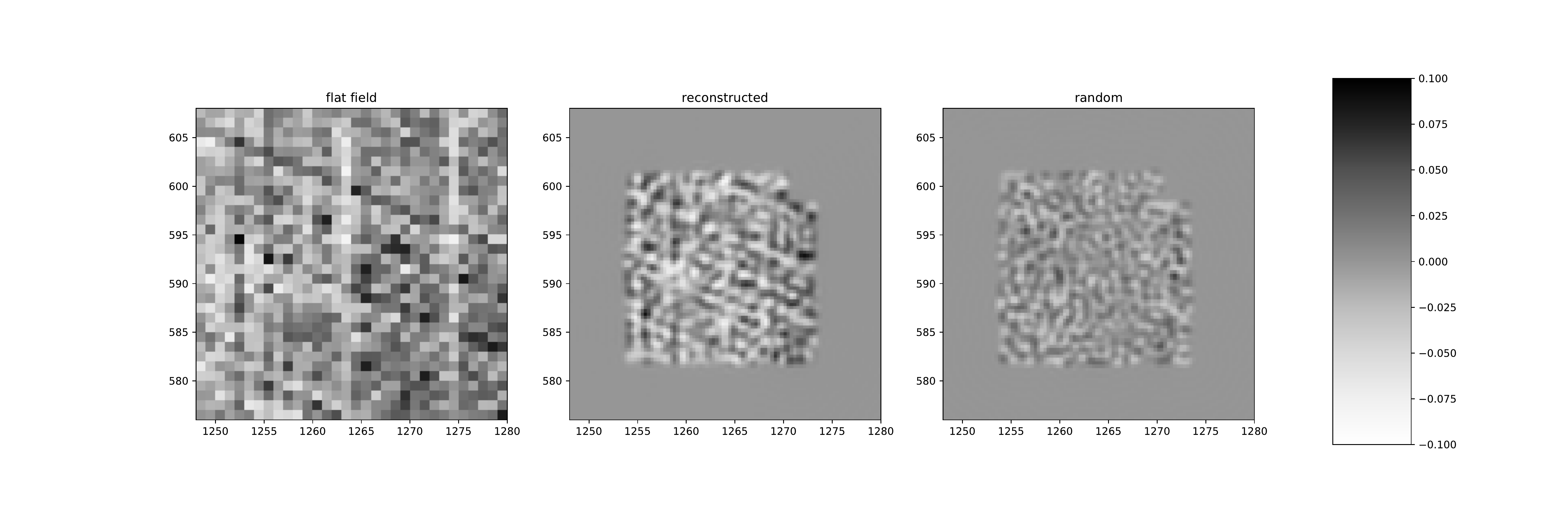}\\
   \end{center}
   \caption{\label{fig:reconstruction_bad} Subpixel reconstructions from a strongly crosshatched detector subregion. Axes units are pixels {\bf Left:} Patch of the flat field image on which this reconstruction was performed.  {\bf Center:} Subpixel response pattern reconstructed from the spot photometry on 1/3 pixel subgrid. {\bf Right:} Same reconstruction procedure, but randomly scrambling the photometry and positions of the spots. This provides a representation of the correlated noise in the reconsruction process.}
 \end{figure}

\begin{figure} [htp]
  \begin{center}
   \includegraphics[width=\textwidth]{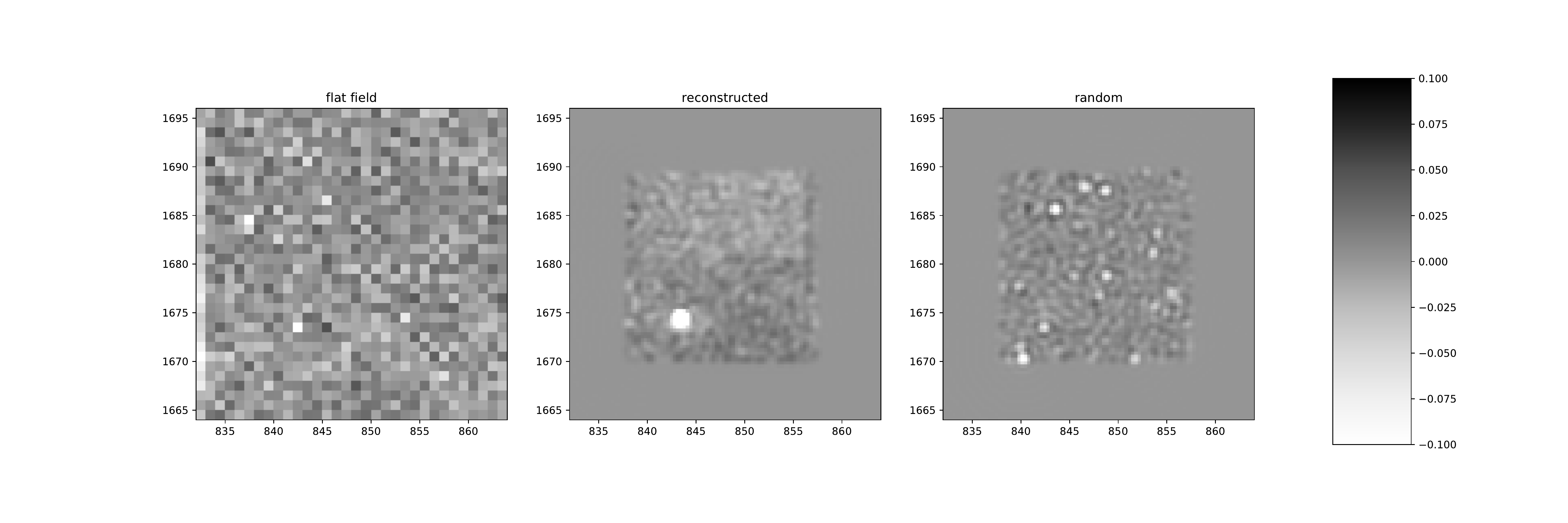}
   \end{center}
   \caption{\label{fig:reconstruction_defect} Same as for figure~\ref{fig:reconstruction_bad}, but for a weak crosshatch region. This region contains a single anomalously low-responsivity pixel which, despite our flat-field correction, retains substantial structure visible in the sub-pixel reponse map.}
   \end{figure}

\begin{figure} [htp]
  \begin{center}
   \includegraphics[width=\textwidth]{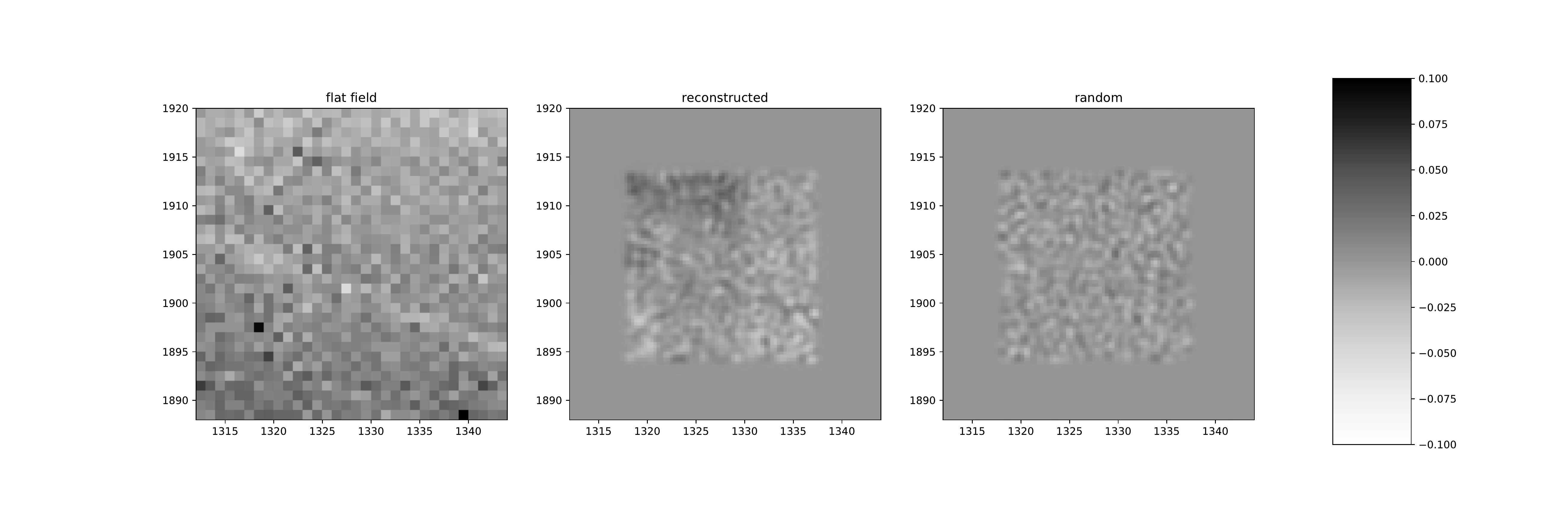}
   \end{center}
   \caption{\label{fig:reconstruction_good} Same as for figure~\ref{fig:reconstruction_bad}, but for a weak crosshatch region. The crosshatching feature visible in the flat at left is mostly removed by flatfielding during data reduction.}
   \end{figure}

We match spots across epochs using their centroid positions, shifted by the mask motor positions mapped onto the pixel coordinates. We subdivide the full detector into $32\times 32$ pixel$^2$ subregions for which we interpolate the photometry measurements onto a grid with a pitch of 1/3 pixels (6 $\um$).  Examples of interpolated subregions are shown in figures~\ref{fig:reconstruction_bad}, \ref{fig:reconstruction_defect}, and \ref{fig:reconstruction_good} alongside images of the corresponding patches of the flat field image.  Note that we correct all images by the flat field and other calibrations \emph{before} measuring photometry, so the reconstructed maps are showing photometric variations that remain \emph{despite} these corrections.

The reconstructed features are often subtle, and (especially in combination with the mildly irregular and occasionally gappy sampling of the subpixel grid) the reconstruction procedure results in correlated noise. We construct realizations of this noise for direct comparison by scrambling the lists of spot photometry and spot positions with respect to one another, and then repeating the reconstruction procedure on this scrambled catalog. These noise realizations are shown in the right panels of figures~\ref{fig:reconstruction_bad}, \ref{fig:reconstruction_defect}, and \ref{fig:reconstruction_good}; it is apparent that the most distinctive crosshatching features are not produced by the correlated noise inherent in the subpixel reconstruction.
Also notable in some reconstructed regions are rectangular-shaped discontinuities (e.g. \reffig{reconstruction_good}). These are subregions reconstructed from two or more different spots in the mask with overlapping scan regions; due to small imperfections in the machining process, the mask apertures are not all the same size. The variations in mean spot intensity mimic variations in the pixel response, but since these are quite distinct from the crosshatching features, we do not attempt any correction in this work.

Our spot map reconstruction procedure clearly reveals residual power that is not removed by flat-fielding the spot images. To further quantify its significance, we compute the 2D mean power spectra of the reconstructed response maps, and subtract from these the 2D power spectra of the scrambled reconstructions. This latter step removes the additive power associated with correlated noise and (partially) corrects for the irregular reconstruction geometry.
There are distinct features in the power spectra of the entire flat field image and the sub-pixel reconstructions which do not pass through the origin at $k=0$. We argue that these features are related: they are largely composed of sub-pixel power which is aliased to lower frequencies where it is detectable in the full-image flat field power spectrum. The operation of dividing the spot images by the response map shifts these features, resulting in a displacement of the aliased subpixel power in the reconstructed power spectra.

\begin{figure} [htp]
  \begin{center}
   \includegraphics[width=\textwidth]{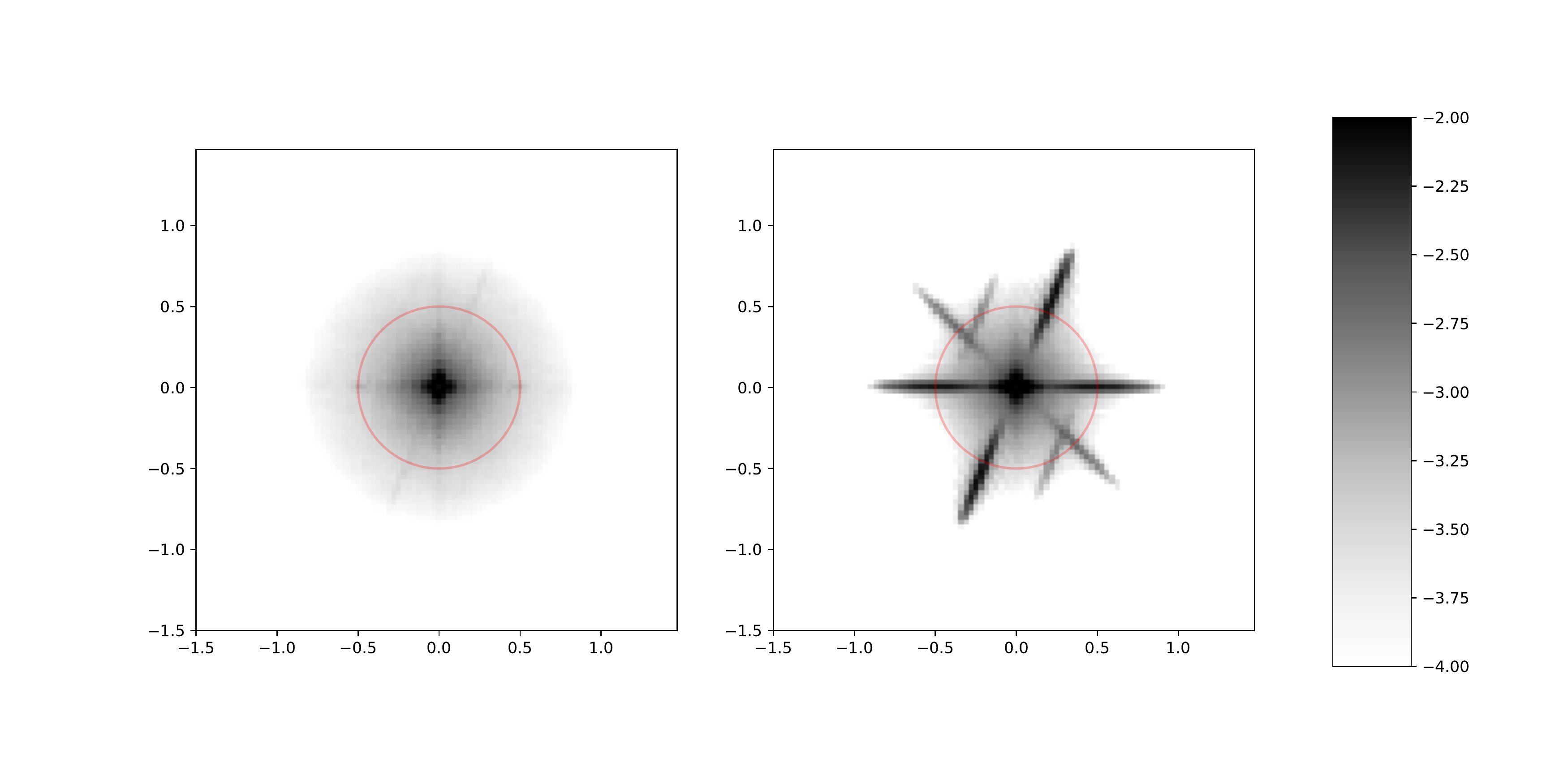}
   \end{center}
   \caption{\label{fig:subpix_power} Noise-subtracted power spectra of subpixel response reconstructions averaged over weak (left) and strong (right) crosshatched regions. Axes are wavenumber, in units of ${\rm pixels}^{-1}$; the pixel Nyquist frequency is indicated with a red circle. Crosshatching signatures along with folded power spikes are clearly visible in the right panel. Faint signatures of the strongest of these features are visible on the left. Color bar units are $\log_{10}P(k)$. 
}
   \label{fig:powerspec}
 \end{figure} 

We test the robustness of our interpretation by simulating a detector with crosshatched flat fields but no sub-pixel structure. 
We simulate erroneous flat fielding by incorporating uncorrected nonlinearity in the pixel response.  
This test represents a case where the sharp edges in the mis-specified flat field pattern produce high-frequency power that might be mistaken for sub-pixel crosshatching.  
We generate crosshatched flat fields with no sub-pixel structure, correct by an arcsinh-rescaled version of the same field, convolve the result with a model PSF and sample from the resulting map to make simulated spot photometry, which is then passed through the same Wiener filter reconstruction code used on our data. 
In this case, even with an unrealistically large uncorrected nonlinearity ($5\%$ of the highest flux level in the flat), nonlinear flat field mis-specification produces primary spikes (those which pass through $k_x=k_y=0$), but no high-frequency aliased power (the offset spikes seen in Figure~\ref{fig:powerspec}). 
We conclude that even fairly gross errors in flat-fielding the spot data cannot explain these features without also invoking some sub-pixel variation in the flat field response.


\section{Conclusions}
We have used the Precision Projector Laboratory to examine the infamous ``crosshatch'' feature seen in the flat-field response pattern of a Teledyne H2RG HgCdTe array. These features are not completely removed during detrending and are detectable as systematic spatial variations in the photometry of spots projected in focus onto the detector.
We use these spot variations to make maps of sub-pixel structure in weakly and strongly crosshatched regions of the H2RG array, finding evidence in our maps and their corresponding power spectra that the crosshatched features are produced by spatial structures in the detector response that are smaller than the pixel scale. 
We propose using the power spectrum of the raw flat-field, which produces a tell-tale signature of the pattern, as a diagnostic when selecting arrays for future applications.  
IR astronomy projects with strict tolerances on detector-induced errors in photometry, astrometry, spectroscopy, or shape measurement will benefit from avoiding detectors with a pronounced crosshatch pattern, especially if the PSF is undersampled.
Where the pattern persists, lab or on-sky calibration with representative sources is needed to fully characterize the pixel response.

   \begin{figure} [htp]
   \begin{center}
   \begin{tabular}{c} 
   \includegraphics[width=5in]{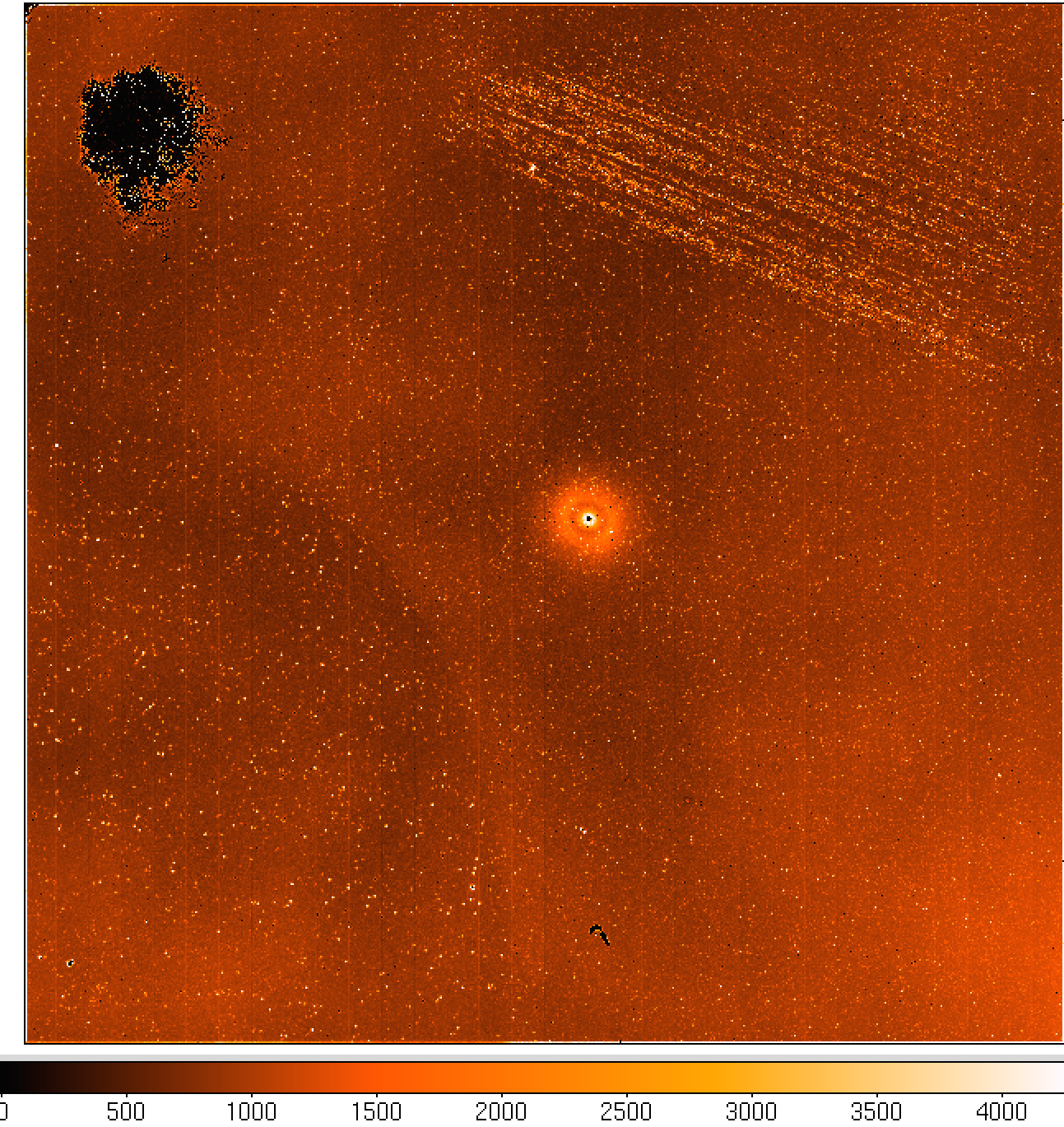}
   \end{tabular}
   \end{center}
   \caption[example] 
   { \label{fig:hotpixel} 
Dark exposure taken when the H2RG was at a temperature of 166K (exposure time 1s).  Color scale units are ADU (gain = 2.7 e-/ADU).  A concentration of hot pixels appears to trace out part of the crosshatch pattern in the weak crosshatch region (compare to \reffig{crosshatch}). }
   \end{figure} 
\appendix
\section{Hot Pixel Anomaly} \label{sec:hotpixel}
While cooling the H2RG to 95K, we acquired diagnostic images to verify normal operation.  
At 166K, we obtained the dark exposure in \reffig{hotpixel}, which displays high dark current and numerous hot pixels as expected for a NIR detector.  
Intriguingly, in the upper right corner, there is a conspicuous concentration of hot pixels resembling the 68 degree component of the crosshatch pattern.  
More intriguingly, this concentration occurs in the \emph{weak} crosshatch region rather than the strong region.  
Close inspection of \reffig{crosshatch} and \reffig{crosshatch-power} shows that the 68 degree component is the most prominent of the three hash directions in the weak region.  
Once the detector reaches its operating temperature, the dark current and hot pixel intensities are greatly attenuated.  
We did not analyze the hot pixel distribution of the detector further since it was outside the scope of our main objectives.  
We present our observation here for completeness and in the hope that it may provide some clues as to the nature of the crosshatch feature and potential mitigations of it.

\acknowledgments 
Thanks to Warren Holmes and the Euclid detector working group for lending the H2RG detector used in this research. 
Thanks to Dave Hale and Alex Delacroix for helping to get the detector up and running.
Thanks also to Mike Seiffert, Stefanie Wachter, and Chris Hirata for valuable insights and critiques of this research.
This work was carried out at the Jet Propulsion Laboratory, California Institute of Technology, under a contract with the National Aeronautics and Space Administration.
Facility space and support for PPL is provided by Caltech Optical Observatories.


\bibliography{references} 

\begin{thebibliography}{10}

\bibitem{Beletic2008}
{Beletic}, J.~W., {Blank}, R., {Gulbransen}, D., {Lee}, D., {Loose}, M.,
  {Piquette}, E.~C., {Sprafke}, T., {Tennant}, W.~E., {Zandian}, M., and
  {Zino}, J., ``{Teledyne Imaging Sensors: infrared imaging technologies for
  astronomy and civil space},'' in [{\em High Energy, Optical, and Infrared
  Detectors for Astronomy III}{\nolinebreak\hspace{0.1em}]},  {\em \procspie}
  {\bf 7021},  70210H (July 2008).

\bibitem{Shapiro2018}
{Shapiro}, C., {Smith}, R., {Huff}, E., {Plazas}, A.~A., {Rhodes}, J., {Fucik},
  J., {Goodsall}, T., {Massey}, R., {Rowe}, B., and {Seshadri}, S.,
  ``{Precision Projector Laboratory: Detector Characterization with an
  Astronomical Emulation Testbed},'' {\em ArXiv e-prints}~{\bf
  arXiv:1801.06599} (Jan. 2018).

\bibitem{Barron2007}
{Barron}, N., {Borysow}, M., {Beyerlein}, K., {Brown}, M., {Lorenzon}, W.,
  {Schubnell}, M., {Tarl{\'e}}, G., {Tomasch}, A., and {Weaverdyck}, C.,
  ``{Subpixel Response Measurement of Near-Infrared Detectors},'' {\em
  \pasp}~{\bf 119},  466--475 (Apr. 2007).

\bibitem{Hardy2014}
{Hardy}, T., {Willot}, C., and {Pazder}, J., ``{Intra-pixel response of the new
  JWST infrared detector arrays},'' in [{\em High Energy, Optical, and Infrared
  Detectors for Astronomy VI}{\nolinebreak\hspace{0.1em}]},  {\em \procspie}
  {\bf 9154},  91542D (July 2014).

\bibitem{Hardy2008}
{Hardy}, T., {Baril}, M.~R., {Pazder}, J., and {Stilburn}, J.~S.,
  ``{Intra-pixel response of infrared detector arrays for JWST},'' in [{\em
  High Energy, Optical, and Infrared Detectors for Astronomy
  III}{\nolinebreak\hspace{0.1em}]},  {\em \procspie} {\bf 7021},  70212B (July
  2008).

\bibitem{Toyozumi2005}
{Toyozumi}, H. and {Ashley}, M.~C.~B., ``{Intra-Pixel Sensitivity Variation and
  Charge Transfer Inefficiency - Results of CCD Scans},'' {\em \pasa}~{\bf 22},
   257--266 (Aug. 2005).

\bibitem{Crouzier2014}
{Crouzier}, A., {Malbet}, F., {Preis}, O., {Henault}, F., {Kern}, P., {Martin},
  G., {Feautrier}, P., {Stadler}, E., {Lafrasse}, S., {Delboulbe}, A., {Behar},
  E., {Saint-Pe}, M., {Dupont}, J., {Potin}, S., {Cara}, C., {Donati}, M.,
  {Doumayrou}, E., {Lagage}, P.~O., {L{\'e}ger}, A., {Le Duigou}, J.~M.,
  {Shao}, M., and {Goullioud}, R., ``{Metrology calibration and very high
  accuracy centroiding with the NEAT testbed},'' in [{\em Space Telescopes and
  Instrumentation 2014: Optical, Infrared, and Millimeter
  Wave}{\nolinebreak\hspace{0.1em}]},  {\em \procspie} {\bf 9143},  91434S
  (Aug. 2014).

\bibitem{Crouzier2016}
{Crouzier}, A., {Malbet}, F., {H{\'e}nault}, F., {L{\'e}ger}, A., {Cara}, C.,
  {Le Duigou}, J.~M., {Preis}, O., {Kern}, P., {Delboulbe}, A., {Martin}, G.,
  {Feautrier}, P., {Stadler}, E., {Lafrasse}, S., {Rochat}, S., {Ketchazo}, C.,
  {Donati}, M., {Doumayrou}, E., {Lagage}, P.~O., {Shao}, M., {Goullioud}, R.,
  {Nemati}, B., {Zhai}, C., {Behar}, E., {Potin}, S., {Saint-Pe}, M., and
  {Dupont}, J., ``{The latest results from DICE (Detector Interferometric
  Calibration Experiment)},'' in [{\em Space Telescopes and Instrumentation
  2016: Optical, Infrared, and Millimeter Wave}{\nolinebreak\hspace{0.1em}]},
  {\em \procspie} {\bf 9904},  99045G (July 2016).

\bibitem{Bertin1996}
{Bertin}, E. and {Arnouts}, S., ``{SExtractor: Software for source
  extraction.},'' {\em \aaps}~{\bf 117},  393--404 (June 1996).

\bibitem{Regan2012}
{Regan}, M., {Bergeron}, E., {Lindsay}, K., and {Anderson}, R., ``{Count rate
  nonlinearity in near infrared detectors: inverse persistence},'' in [{\em
  Space Telescopes and Instrumentation 2012: Optical, Infrared, and Millimeter
  Wave}{\nolinebreak\hspace{0.1em}]},  {\em \procspie} {\bf 8442},  84424W
  (Sept. 2012).

\bibitem{Plazas2014}
{Plazas}, A.~A., {Bernstein}, G.~M., and {Sheldon}, E.~S., ``{On-Sky
  Measurements of the Transverse Electric Fields' Effects in the Dark Energy
  Camera CCDs},'' {\em \pasp}~{\bf 126},  750 (Aug. 2014).

\bibitem{1992ApJ...398..169R}
{Rybicki}, G.~B. and {Press}, W.~H., ``{Interpolation, realization, and
  reconstruction of noisy, irregularly sampled data},'' {\em ApJ}~{\bf 398},
  169--176 (Oct. 1992).

\end{thebibliography}
\bibliographystyle{spie-proceedings-style/spiebib} 

\end{document}